# Comprehensive understanding of size-, shape-, and composition-dependent polarizabilities of $Si_mC_n$ ($m$, $n$ = 1–4) clusters

You-Zhao Lan, Hong-Lan Kang, Tao Niu

Institute of Physical Chemistry, Zhejiang Normal University, Jinhua 321004, China

Correspondence to: You-Zhao Lan (E-mail: *lyzhao@zjnu.cn*)

**Abstract**

We performed a comprehensive study of the size-, shape-, and composition-dependent polarizabilities of $Si_mC_n$ ($m$, $n$ = 1–4) clusters on the basis of the density-functional-based coupled perturbed Hartree-Fock calculations. We found better correlations between the polarizabilities and both the binding energies ($E_b$) and change in charge distribution ($\Delta q$) than the energy gaps ($E_g$). The α values exhibit overall decreasing and increasing trends with increases in the $E_b$ and $\Delta q$ values, respectively. For isomers with the same $E_b$ values and different polarizabilities, $\Delta q$ can well explain the difference in polarizabilities. The π-electron delocalization effect is the best factor for understanding the shape-dependence. For a given $m/n$ value, the linear clusters have an obviously larger polarizability than both the prolate and compact clusters, irrespective of the cluster size. We fit a quantitative expression [α = A – (A – B) × exp(–k(m/n))] to describe the composition-dependent polarizabilities.

## 1. Introduction

In the last 20 years, the (hyper)polarizabilities of small semiconductor clusters such as gallium arsenide (GaAs), silicon (Si), silicon carbide (SiC), and aluminum phosphide (AlP) clusters have attracted much attention [1–27]. Experimental studies have shown that the static polarizabilities of $Si_m$ ($m$ = 9–50) and $Ga_mAs_n$ ($m + n$ = 5–30) clusters fluctuate around their corresponding bulk values [12]. These experimental results have motivated many theoretical studies of the polarizabilities of Si and GaAs clusters [1, 2, 4, 6 – 9, 16, 26, 27]. Theoretical studies have shown that the size of the polarizability directly or indirectly depends on various factors such as the cluster size, cluster shape, cluster composition, energy gap, binding energy, ionization potential, and so on. The size-dependence of the polarizabilities has been well known for Si and GaAs clusters. For small $Si_m$ ($m$ < 10) and $Ga_nAs_m$ ($n + m$ < 8) clusters, the theoretical polarizabilities are higher than the bulk value and decrease with an increase in cluster size [1, 2, 9, 12, 16], as indicated by the number of atoms in a cluster. For mediate-size $Si_m$ ($m$ = 9–50) and $Ga_nAs_m$ ($n + m$ = 5–30) clusters, the experimental polarizabilities [12] vary strongly and irregularly with the cluster size and fluctuate around the bulk value. For large $Si_m$ ($m$ = 60–120) clusters, all the experimental polarizabilties are lower than the bulk value [12]. However, for $Si_m$ ($m$ = 9–28) clusters, Deng *et al.* [6] found that the theoretical polarizabilities exhibit fairly irregular variations with the cluster size, and all the calculated values are higher than the bulk value. Similar theoretical results have also been obtained by Sieck *et al.* [21] and Jackson *et al.* [7] for $Si_m$ ($m$ = 1, 3–14, 20, and 21) and ($m$ = 1–21) clusters, respectively. Until now, the discrepancies between the experimental and theoretical polarizabilities have not been well explained [6,7]. More investigation is required.

The shape-dependence of the polarizabilities has also been known for Si, GaAs, and AlP clusters [4, 7, 23, 28]. For $Si_m$ ($m$ = 20–28) clusters, the prolate clusters have a systematically larger polarizability than the compact ones [4]. This shape-dependence reflects the metallic character of these $Si_m$ ($m$ = 20–28) clusters, because it is reproduced by using the jellium models. For the prolate $(GaAs)_n$ and $(AlP)_n$ clusters [22, 23, 28], the (hyper)polarizabilities monotonically increase with an increase in $n$. This trend is very similar to the well known (hyper)polarizability evolution of extended conjugated organic molecules [29]. We also notice other interesting shape-dependences of the polarizabilities. For $Co_n(C_6H_6)_m$ ($n, m$ = 1–4, $m = n, n + 1$) clusters [30], the sandwich clusters have systematically larger polarizabilities and anisotropies than the rice-ball isomers, which suggests that we should distinguish these two kinds of clusters in terms of their dipole polarizabilities. The (hyper)polarizabilities of the Möbius, normal cyclacene, and linear nitrogen-substituted strip polyacenes exhibit clear shape-dependence [31]. The shape-dependence based on the geometries of the cyclacene with and without a knot, namely, Möbius and normal cyclacene, is very different from that based on the linear, prolate, and compact geometries of semiconductor clusters. Therefore, more different shape-dependences could be considered for semiconductor clusters in the future. The composition-dependence of the polarizabilities is only available for heteroatomic clusters and is less known than the size- and shape-dependences for semiconductor clusters [12, 26]. Karamanis *et al.* [26] investigated the composition-dependent polarizabilities of open- and closed-shell $Ga_mAs_n$ clusters with $m + n$ = 5 and 6. They showed that for a given size (5 or 6), the polarizabilites of the $Ga_mAs_n$ clusters gradually increase with an increase in the number of Ga atoms in a cluster. This dependence implies that for the heteroatomic clusters, we can obtain a tunable polarizability by adjusting the composition of the clusters.

Understanding the evolution of the polarizability is essential for nanomaterial design. For example, on the basis of the composition-dependence of the polarizability, we have more choices to obtain different polarizabilities by adjusting cluster composition. Although the size-dependence has generally been correlated with the energy gaps ($E_g$) between the highest occupied molecular orbital (HOMO) and lowest unoccupied molecular orbital (LUMO), many studies have shown that the correlation between $\alpha$ and $E_g$ is very poor [4, 6–8]. Therefore, it is necessary and interesting to seek other factors to understand the size-dependent polarizabilities. In this work, we find better correlations between the polarizabilities and binding energies ($E_b$) than $E_g$. For the isomers with the same $E_b$ values and different $\alpha$ values, we show that the change in the charge distribution ($\Delta q$) can be used to understand the size of the polarizability. Meanwhile, to explicitly describe the composition-dependent polarizabilities, we fit a quantitative expression for the relationship between the $\alpha$ values and the cluster composition such as $m/n$ in the $Si_mC_n$ clusters.

In Section 2, we provide the computational details. In Section 3, we discuss the size-, shape-,

and composition-dependent polarizabilities of $Si_mC_n$ ($m$, $n$ = 1–4) clusters. Finally, we summarize our results in Section 4.

## 2. Computational Details

There have been many studies on the geometry and electronic structure of small SiC clusters. Low-lying isomers of these clusters have been theoretically or experimentally determined. For the purpose of our present work, we selected the lowest-lying isomers and some low-lying isomers with different shapes (*i.e.*, linear (chain), prolate (flat), and compact clusters) of the $Si_mC_n$ ($m$, $n$ = 1–4) clusters on the basis of the literature [32–35]. Note that for the small clusters considered here, the linear and prolate isomers have linear chain and flat forms, respectively, which is different from the classification for the larger clusters reported in previous studies. Although a linear geometry can be considered to be prolate, for the small clusters of less than ten atoms considered in this work, linear clusters have obviously different polarizabilities from the prolate clusters. Therefore, these linear clusters were classified individually. For a given shape, the selected low-lying isomers had the lowest energy. For example, the linear $D_{\infty h}$ structure was the lowest-lying isomer or ground-state structure of $Si_2C_3$, whereas the $C_{2v}$ prolate structure was the lowest energy isomer for the prolate $Si_2C_3$ clusters [32]. Note that there have been some discrepancies between the results of theoretical and experimental studies on the ground-state structure of SiC clusters (even small clusters). For instance, most theoretical and experimental studies [36–39] have revealed that $SiC_3$ has three stable isomers (two having four-membered rings ($C_{2v}$, $^1A_1$) and one with a linear structure ($C_{\infty v}$, $^3\Sigma$)) and that the global minimum structure of $SiC_3$ is a $C_{2v}$ prolate with a transannular C–C bond. However, these results are still controversial because the linear structure was predicted to be the ground state structure on the basis of highly accurate coupled cluster methods, including a perturbative treatment of triple excitations and Dunning's correlation-consistent polarized core-valence quadruple zeta basis set [CCSD(T)/cc-pCVQZ] [40]. In the $Si_2C$ cluster [32, 33], the $C_{2v}$ and $D_{\infty h}$ singlet structures compete for the ground-state structure, depending on the level of theory. All of the selected clusters were re-optimized at the DFT/aug-cc-pVTZ level using the hybrid Becke3-Lee-Yang-Parr (B3LYP) functional. The vibrational frequencies were calculated to confirm that the final geometries are stable without an imaginary frequency. The final geometries and their electronic states, symmetries, cluster shape, binding energies ($E_b$), and energy gaps ($E_g$) are shown in figure 1.

For the estimation of the dipole polarizability ($\alpha$), we focused on the isotropic dipole polarizability ($<\alpha>$), which is defined as $<\alpha> = 1/3(\alpha_{xx} + \alpha_{yy} + \alpha_{zz})$. The $<\alpha>$ is expressed in Å$^3$/atom. Hereafter, $\alpha$ refers to $<\alpha>$. To obtain accurate dipole polarizabilities, we should select an appropriate theoretical method and reasonable basis set. In general, the diffuse and polarization functions and electron correlation effects should be considered in the calculation. For density-functional-based methods, the local density approximation (LDA) and general gradient approximation (GGA) functionals can produce the close dipole polarizabilities. For example, the Vosko-Wilk-Nusair (VWN), BLYP, and B3LYP functionals produce polarizabilities of 5.17, 5.52, and 5.37 Å$^3$/atom for the $C_{2v}$ singlet $Si_3$ cluster [8], respectively. Different hybrid functionals such as the B3LYP, B3P86, and B3PW91 functionals also produce very close polarizabilities [16, 22]. For wave-function-based methods, the Møller-Plesset second order perturbation theory (MP2) with the aug-cc-pVDZ or 6-31G augmented by the standard diffuse and polarization functions can lead to accurate dipole polarizabilities, which are very close to the results based on highly accurate coupled cluster singles-and-doubles calculations, including a perturbative triples correction for binary semiconductor clusters such as AlP and GaAs clusters [22–24, 26]. A more detailed discussion of the basis-set and theoretical method dependences of $\alpha$ for small Si and SiC clusters can be found elsewhere [2, 8, 9, 15, 16]. In this work, we calculated the polarizabilities by using the coupled perturbed Hartree–Fock (CPHF) approach at the B3LYP/aug-cc-pVTZ level. A test B3LYP/aug-cc-pVTZ calculation performed on a CO molecule gave a polarizability of 1.95 Å$^3$, which is in good agreement with the experimental [41] and theoretical [42] value of 1.95 Å$^3$. The calculated polarizabilities are also shown in figure 1. The polarizability of the SiC molecule was calculated using both the CPHF and finite field (FF) method at the aug-cc-pVTZ level to make a comparison with previous results [43] and because there was no appropriate symmetry ($\alpha_{xx}$: 57.35, $\alpha_{yy}$: 34.46, and $\alpha_{zz}$: 59.54 a.u.) of the three polarizability components based on the CPHF/B3LYP calculation. The FF/B3LYP and FF/MP2 methods based on the fields of 0.003 (parallel) and 0.001 (perpendicular) a.u. give isotropic polarizabilities of 4.40 and 3.84 Å$^3$/atom, respectively. The FF/MP2//aug-cc-pVTZ result is in agreement with the polarizability of 3.86 Å$^3$/atom based on the FF/MP2 with a self-designed basis set [43] and the fields of 0.003 (parallel) and 0.001 (perpendicular) a.u. The FF/B3LYP/aug-cc-pVTZ result was used for the SiC molecule in the following discussion. All of the calculations were performed using the Gaussian 03 program [44].

## 3. Results and Discussion
### 3.1. Size-dependence

We first investigate the size-dependent polarizabilities of Si$_m$C$_n$ ($m$, $n$ = 1–4) clusters. It can be expected that the size-dependence of polarizabilities of heteroatomic clusters will be more complicated than that of homoatomic clusters because heteroatomic clusters with a given cluster size have more isomers than homoatomic ones. Figure 2 shows the size-dependent polarizabilities of Si$_m$C$_n$ ($m$, $n$ = 1–4) clusters. The cluster size is indicated by the number of atoms ($N$) in cluster. For comparison, the polarizabilities based on the B3LYP calculations of Si$_m$ ($m$ = 1–8) (Ref. [8, 16]), and C$_n$ (linear: $n$ = 1–8; cyclic: $n$ = 4, 6, and 8) (Ref. [45]) clusters are included in figure 2. In Ref. [45], the $\alpha$ values of the C$_n$ ($n$ = 1–8) clusters were calculated using a self-designed basis set that is different from the aug-cc-pVTZ we used here. Therefore, we recalculated the $\alpha$ values of C$_n$ ($n$ = 1–8) clusters to check the difference caused by the basis set effect. The obtained results are close to those of Ref. [45] and the difference is lower than ~0.03 Å$^3$/atom. For example, we obtained a polarizability of 1.86 Å$^3$ for C atom, which is close to 1.88 Å$^3$ reported in Ref. [45]. The Si, SiC, and C bulk values are also included in figure 2 and they are 3.71 (Ref. [1]), 1.84 (Ref. [15]), and 0.83 Å$^3$/atom, respectively, which were obtained according to the Clausius-Mossotti relation. We calculated the C bulk value on the basis of the dielectric constant of 5.7 and the diamond density of 3.51 g/cm$^3$. As expected, small SiC clusters have a more complicated size-dependence of $\alpha$ than small Si and C clusters. For small Si$_m$ ($m$ < 10) clusters, the $\alpha$ values are higher than the corresponding bulk value and decrease with an increase in cluster size [1, 8, 12, 16, 45]. For small *linear* C$_n$ ($n \leq 8$, $n \neq 2$) clusters, we can observe an increasing trend with an increase in cluster size, which is attributed to an increasing trend of longitudinal component of $\alpha$. For small SiC clusters, we cannot observe a simple trend as small Si and C clusters. As shown in figure 2, the $\alpha$ values of Si$_m$C$_n$ ($m$, $n$ = 1–4) clusters are higher than the corresponding bulk value and have an oscillating variation with the cluster size.

We can also see from figure 2 that the $\alpha$ values of the SiC clusters lie between those of pure Si and C clusters (*i.e.*, an order of $\alpha$(Si) > $\alpha$(SiC) > $\alpha$(C)), especially when the cluster size is fixed. For example, for diatomic clusters, $Si_2$, SiC, and $C_2$ clusters, their polarizabilities are 7.35, 4.30, and 3.63 Å$^3$/atom, respectively. The same case occurs for triatomic clusters. For some given cluster sizes, this order does not hold and the SiC cluster has a larger $\alpha$ than the Si cluster or a smaller $\alpha$ than the C cluster. For example, for $n$ = 6, the linear $Si_2C_4$ cluster has a larger $\alpha$ of 5.07 Å$^3$/atom (figure 1) than the compact $Si_6$ cluster with 4.56 Å$^3$/atom. For $n$ = 8, the linear $C_8$ cluster has a larger $\alpha$ of 3.13 Å$^3$/atom than the prolate $C_s$(b) $Si_4C_4$ cluster with 2.94 Å$^3$/atom (figure 1). These exceptions exist because for these clusters the shape effect (*i.e.*, linear π delocalized structure) significantly determines the size of $\alpha$ (see discussion below). For $N$ > 10, a comparison can not be made because of lack of the polarizabilities of the SiC clusters.

To understand the size of $\alpha$ of the clusters, we first consider the approach based on the energy gap. In the sum-over-states (SOS) expression of polarizability obtained from the simple perturbation theory,

$$\alpha_{ii} = 2\sum\nolimits_{k,l}' \frac{\left|\langle k|\mu_i|l\rangle\right|^2}{E_l - E_k},$$

where the matrix element $<k|\mu_i|l>$ corresponds to the transition dipole moment between occupied ($l$) and unoccupied ($k$) states in the $i$th direction, and $E_l - E_k$ is the corresponding transition energy, if we assume that the transitions between the HOMO and the LUMO make a major contribution to the polarizability, then the $\alpha$ can be inversely related to the $E_g$ (Refs. [1, 7, 8]). However, the inverse relationship between $\alpha$ and $E_g$ does not hold in general. For $Si_m$ ($m$ < 30) clusters, theoretical researches have shown that the correlation between $\alpha$ and $E_g$ is very weak [4, 6, 7]. For example, for $Si_m$ ($m$ = 11–28) clusters, the overall distribution in the plot of $\alpha$ versus $E_g$ is very scattered [6]. However, a research on the germanium clusters with 2 – 25 atoms [46] have shown a well correlation between $\alpha$ and $E_g$, that is, the size of $\alpha$ is inversely related to the size of $E_g$ and a molecule with a smaller $E_g$ is found to be softer and has a larger $\alpha$. To understand the lack of correlation between $E_g$ and $\alpha$, Pouchan *et al.* [8] have shown that this lack can be understood as the vanishing matrix element $|<k|\mu_i|l>|^2$ between HOMO and LUMO in the SOS expression for some clusters, especially for the cluster with small size and high symmetry, and they have made a direct correlation between the $\alpha$ and the lowest symmetry-allowed transition energy gap not the energy gap for small $Si_m$ ($m$ = 3–10) clusters on the basis of the density function calculations with different functionals. In figure 3, we plot the $\alpha$ values versus the $E_g$ values (figure 1) for $Si_mC_n$ ($m$, $n$ = 1–4) clusters. The results reveal an irregular correlation between $\alpha$ and $E_g$ and a very scattered overall distribution.

On the other hand, one correlates the size of $\alpha$ with $E_g$ on the basis of the inverse relationship between $\alpha$ and hardness. A harder molecule [47] has a larger $E_g$, in other words, a stable system with a large $E_g$ (Ref. [48]) has a low polarizability according to the minimum polarizability principle [49–51] which points out that "the natural direction of evolution of any system is towards a state of minimum polarizability". Note that the stability of a system can be also related to its $E_b$ value [48], that is, a stable system has a large $E_b$. Therefore, we secondly attempt to correlate $\alpha$ with $E_b$. Figure 4 shows the plots of $\alpha$ versus $E_b$ for $Si_mC_n$ ($m$, $n$ = 1 – 4) clusters. Compared with $E_g$, the $\alpha$ values have an overall decreasing trend with an increase of the $E_b$ values, which is approximately indicated by a dot line in figure 4. In the following discussions, we will find a better correlation between $\alpha$ and $E_b$ when both the cluster shape and one of the components in cluster are fixed.

Thirdly, we correlate the size of $\alpha$ with the change in charge distribution related to an external field. Jackson *et al.* [4] used this approach to reveal that the response of the compact and prolate $Si_m$ ($m = 20 - 28$) clusters to a static external field is metallic on the basis of the metallic-like distribution of charge in these clusters. Karamanis *et al.* [22] employed this approach to show that the bonding effect is more important than the cluster composition on the hyperpolarizablity values for a series of $Al_nP_n$ ($n = 2, 3, 4, 6,$ and 9) clusters. The response of a molecule to an external field leads to a redistribution of atomic charge in cluster, especially for the surface atoms [4]. In our present work, we used a modified parameter:

$$\Delta q = \frac{1}{3N}\sum_j \Delta q_j = \frac{1}{3N}\sum_j \sum_i^N \left| q_i(F_j) - q_i(0) \right|$$

to characterize the redistribution of charge, where $N$ is the cluster size, $q_i(F_j)$ is the natural charge of atom $i$ perturbed by a $F_j$ external field in the direction $j = x, y, z$, and $q_i(0)$ is the natural charge of an unperturbed molecule. To obtain distinct $\Delta q$ values, we applied a strong field of 0.02 a.u. to calculate the natural charge. We collected the calculated results in table 1 where three polarizability components ($\alpha_{xx}$, $\alpha_{yy}$, and $\alpha_{zz}$) are also included. From table 1, we observe the same size order for the components of $\alpha$ and $\Delta q$. For example, the prolate $SiC_2$ cluster has orders of $\alpha_{xx} < \alpha_{yy} < \alpha_{zz}$ and $\Delta q_x < \Delta q_y < \Delta q_z$. We assume that the cluster with a larger $\Delta q$ would have a larger polarizability because the electric polarization leads the charge distribution of a molecule to distort from its normal shape. Figure 5 shows the plots of $\alpha$ versus $\Delta q$ for the $Si_mC_n$ ($m, n = 1 - 4$) clusters. Compared with the $E_g$, the $\alpha$ values have an overall increasing trend with an increase of the $\Delta q$ values, which is approximately indicated by a dot line in figure 5. Similar to the $E_b$, we will find a better correlation between $\alpha$ and $\Delta q$ provided that both the cluster shape and one of the components in cluster are fixed (see discussion below).

*3.2. Shape-dependence*

It is well known that organic molecules with a linear or prolate geometry have a large (hyper)polarizability because they have a delocalized π-conjugated structure [29, 52]. Constructing a π-conjugated structure has become one of choices to design a new molecule with large (hyper)polarizability [53]. For small SiC clusters with increasing the cluster size, three alternative hybridizations ($sp$, $sp^2$, and $sp^3$) of C atom result in a variation from linear to prolate, then compact structure in lowest-lying isomers while the $sp^3$ hybridization of Si atom leads to a variation from prolate to compact structure. A transition from prolate to compact will lead to a decrease of the electron delocalization, ultimately, a decrease of the (hyper)polarizability [22, 24]. Experimental and theoretical studies have provided sufficient information on the geometry [32, 33, 54–59] for $Si_mC_n$ ($n + m < 8$) clusters. The structure of the SiC cluster is a result of the competition for bonding that occurs between the C and Si atoms. As shown in figure 1, C-rich clusters tend to exist in the linear or prolate structure while Si-rich clusters prefer forming prolate or compact structure. For example, the lowest-lying isomers of $SiC_n$ ($n = 2, 3,$ and 4) clusters are the linear structure while those of $Si_mC$ ($m = 2, 3,$ and 4) clusters are prolate or compact structure (figure 1). In our recent work [15], we have shown that the size-dependence of the first-order hyperpolarizabilities of $SiC_n$ ($n = 2 - 6$) clusters, which have approximate Si-terminated linear chain geometry, is similar to that observed in π-conjugated organic molecules. For semiconductor clusters such as Si, AlP, and GaAs clusters, theoretical researches have shown that the prolate clusters have systematically larger polarizabilities than the compact ones [4, 6, 7, 22, 23, 27, 28]. Therefore, the size of the (hyper)polarizability, to some degree, depends on the geometry or shape of a cluster.

Figure 6 shows the shape-dependence of $\alpha$ for $Si_mC_n$ ($m$, $n$ = 1–4) clusters. Although no clear correlation is observed between the size of $\alpha$ and the cluster shape from figure 6, we can see that most of the linear clusters have a large $\alpha$ and the shape-dependence of $\alpha$ will be clear when the cluster size and composition are fixed. For example, for triatomic clusters ($Si_2C$ and $SiC_2$), the linear and prolate $Si_2C$ clusters have the $\alpha$ values of 4.18 and 4.03 Å$^3$/atom, respectively, and for the linear and prolate $SiC_2$ clusters they are 2.73 and 2.25 Å$^3$/atom, respectively (figure 1). A similar case occurs for tetra-atomic clusters ($Si_2C_2$ and $SiC_3$ clusters). In detail, figure 7a shows the plots of the $\alpha$ values versus the number of C atoms ($n$) in cluster with both the cluster shape and the number of Si atoms fixed. From figure 7a, we can clearly observe orders of $\alpha(L) > \alpha(P)$ and $\alpha(P) > \alpha(C)$ for a given composition. For the prolate and compact series clusters, the $\alpha$ values decrease with an increase of the number of C atoms ($n$) in cluster, which indicates a composition-dependence of $\alpha$ (see discussion below).

To understand the shape-dependence of $\alpha$ for the $Si_mC_n$ ($m$, $n$ = 1 – 4) clusters, we attempt to use the following approaches:

(1) Using both $E_g$ and $E_b$. As mentioned above, both $E_g$ and $E_b$ have no regular overall variation with the size of $\alpha$. In figures 7b and 7c, we plot the $E_b$ and the $E_g$ versus the number of C atoms ($n$) in cluster, respectively, corresponding to the clusters considered in figure 7a. On the basis of the SOS expression, we would have a larger $\alpha$ for the cluster with a smaller $E_g$, but in figure 7c this behavior has not been observed for all clusters except $SiC_4$ (L and P) and $Si_3C_3$ (P and C). Therefore, the shape dependence cannot be explained by the $E_g$ even when both the cluster size and the cluster shape are fixed, in accord with the Jackson *et al.* [4] who had shown that the differences between the $\alpha$ values of the prolate and compact clusters cannot be explained on the basis of the $E_g$ for $Si_m$ ($m$ = 20–28) clusters. For the $E_b$, for all series of clusters except $Si_2C_n$ (L) we can observe an inverse relationship between $\alpha$ and $E_b$, in agreement with the fact that a cluster has a high stability and a low polarizability.

(2) Using a geometrical parameter,

$$I = \sum_{i=1}^{n} r_i^2,$$

where $r_i$ is the distance of atom $i$ to the cluster center of mass and $n$ is the number of atoms in cluster. Deng *et al.* [6] used this approach to explain the evolution of $\alpha$ for $Si_m$ ($m$ = 9 – 28) clusters and showed that more elongated clusters are more polarizable. However, Jackson *et al.* [4] found that the shape-dependence of the *total* polarizabilities is clearly more complicated than the difference in $I$ between the compact and prolate $Si_m$ ($m$ = 20–28) clusters. For the SiC clusters we considered here, the $I$ values are close between different shape clusters especially for a given cluster size because the clusters have small size ($N \leq 8$). Therefore, this approach is not available for the systems considered here.

(3) Using the $\Delta q$ defined above. In figure 7d, we plot the $\Delta q$ versus the number of C atoms ($n$), corresponding to the clusters considered in figure 7a. Combining with figure 7a, we can see that the shape dependence of $\alpha$ cannot be reflected by the size of the $\Delta q$. For example, for the linear and prolate $Si_2C_n$ clusters with a given $n$, the linear $Si_2C_n$ cluster has a larger $\alpha$ than compact $Si_2C_n$ one (figure 7a) while the size orders of Δ$q$ between prolate and compact clusters alternately change with increasing the number of C atoms ($n$) (figure 7d). For the prolate and compact Si$_3$C$_n$ clusters with a given $n$, reverse size orders are observed for the $α$ and Δ$q$ values. However, it is interesting that the $α$ values have the same variation trend as the Δ$q$ with an increase of the number of C atoms for each series of clusters. Comparing figure 7d with figure 7b, we can see that the Δ$q$ is more available than the $E_b$ for qualitatively understanding the size of $α$. For instance, for both the linear and the prolate Si$_2$C$_n$ clusters, the Δ$q$ values can well reflect the evolutions of $α$ (figure 7a and figure 7d) while the $E_b$ values cannot (figure 7a and figure 7b). Furthermore, some isomers with very close $E_b$ values and different $α$ values may be identified by the Δ$q$ values. For example, for Si$_4$C$_4$(C$_{2v}$), Si$_4$C$_4$(C$_s$(a)), and Si$_4$C$_4$(C$_s$(b)) clusters whose $E_b$ values are 4.486, 4.484, and 4.487 eV/atom (figure 1), respectively, the size of $α$ (Si$_4$C$_4$(C$_{2v}$): 3.46 Å$^3$/atom, Si$_4$C$_4$(C$_s$(a)): 3.36 Å$^3$/atom, Si$_4$C$_4$(C$_s$(b)): 2.94 Å$^3$/atom) is well related to that of Δ$q$ (Si$_4$C$_4$ (C$_{2v}$): 0.125 e/atom, Si$_4$C$_4$ (C$_s$(a)): 0.094 e/atom, and Si$_4$C$_4$ (C$_s$(b)): 0.076 e/atom, respectively).

(4) Considering the delocalization of π-electron structure. Organic molecules with π-electron delocalization have a large (hyper)polarizability [29, 52] because a π-electron delocalization leads to a strong charge separation. As shown in figure 1, both the linear and the prolate clusters would form a π-electron delocalization framework. Although the transition between the HOMO and the LUMO does not determine the size of $α$, the transitions between different frontier MOs generally make significant contributions to the size of $α$ on the basis of the SOS expression [15]. To check the distribution of electron, we provide in figure 8 the HOMO−1, HOMO, LUMO, and LUMO+1 of SiC$_2$, SiC$_4$, Si$_3$C$_2$, Si$_3$C$_3$, and Si$_4$C$_4$ clusters based on the B3LYP/aug-cc-pVTZ wave function. It is clearly shown that for a given cluster size, the linear cluster has more π-delocalized frontier molecular orbital (MO) than the prolate one. For example, the linear SiC$_2$ cluster has four π-delocalized MOs (two-fold degenerate HOMO−1 and LUMO) while the prolate SiC$_2$ cluster has only one π-delocalized MO (LUMO+1). Similarly, for the prolate and compact clusters, the prolate cluster has more obvious π-delocalized MO than the compact one. Therefore, the shape-dependence of $α$ can be well understood on the basis of the electron delocalization of these frontier MOs.

*3.3. Composition-dependence*

As mentioned above, for heteroatomic clusters such as AlP and GaAs clusters, the size of $α$ also strongly depends on the cluster composition [25, 26]. Figure 9 shows the plots of the $α$ values versus the $m/n$ values for Si$_m$C$_n$ ($m$, $n$ = 1–4) clusters. Regardless of the cluster shape, we cannot observe a clear correlation between $α$ and $m/n$ from figure 9. The linear clusters have an obvious larger polarizability than both the prolate and compact clusters especially for a given $m/n$ value. We notice that four linear clusters with small $m/n$ ratio have a large $α$. These four clusters are SiC, Si$_2$C$_4$, Si$_2$C$_3$, and Si$_2$C$_2$ with 4.30, 5.07, 4.26, and 4.75 Å$^3$/atom, respectively. Combining with the shape-dependence of $α$, we can conclude that the shape effect makes a main contribution to the size of $α$ for these four clusters because they have very different cluster sizes ($N$ = 2, 6, 5, and 4, respectively) and small $m/n$ ratios (≤ 1.0). When excluding the linear clusters, we can observe an overall increasing trend of $α$ with an increase of $m/n$. A nonlinear fit for the $α$ values versus the $m/n$ values of the prolate and compact clusters give an expression of $α = A − (A − B) × \exp(−k(m/n))$, where $A$ = 4.3, $B$ = 1.5, and $k$ = 0.87. The $A$ and $B$ values locate at the region of the $α$ values of pure Si and C clusters, respectively (figure 2). This expression indicates that the $α$ values of small SiC clusters would tend towards those of pure $Si_m$ clusters with $n = 0$ and towards those of pure $C_n$ clusters with $m = 0$. This dependence is useful for us to design SiC clusters with tunable $α$ values. Note that the tunable polarizabilities presented here are based on the stable clusters. For unstable or metastable clusters such as transition and excited state clusters, the tunable behavior of polarizabilities is difficult to obtain because the excited state polarizabilities are very different from the ground state ones [60, 61].

Similar to the GaAs cluster [26], we can understand the composition-dependence of $α$ in terms of the atomic polarizabilities of C and Si atoms. On the basis of the B3LYP/aug-cc-pVTZ calculations, we obtained the $α$ values of 1.86 and 6.00 Å$^3$ for C and Si atoms, respectively. The $α$ value of Si atom is three times larger than that of C atom. As shown in the polarizabilities of GaAs clusters [26], a replacement of As by Ga in cluster will increase the polarizability because the $α$ of Ga atom is almost twice that of As atom. A same case occurs for the SiC cluster we considered here provided that the cluster shape is fixed. Actually, all the lowest-lying GaAs clusters considered by Karamanis *et al.* [26] have a compact structure except that the $Ga_4As$ with a prolate structure has a largest $α$ among four pentaatomic GaAs clusters. In our present work, for $N = 5$, the compact $Si_4C$ (4.11 Å$^3$/atom), prolate $Si_3C_2$ (3.85 Å$^3$/atom), and linear $Si_2C_3$ (4.26 Å$^3$/atom) clusters individually have a larger $α$ than the compact $Si_3C_2$ (3.42 Å$^3$/atom), prolate $Si_2C_3$ (3.05 Å$^3$/atom), and linear $SiC_4$ (3.02 Å$^3$/atom) clusters. Note that the linear $Si_2C_3$ cluster has a largest $α$, which further implies that the shape effect makes a main contribution to determining the size of $α$.

From figure 7a, we have seen that the $α$ values decrease with an increase of the number of C atoms ($n$) when both the cluster shape and the number of Si atoms in clusters are fixed. Furthermore, in figure 10, we plot the $α$ values versus the number of Si atoms ($m$) with both the cluster shape and the number of C atoms fixed. All of the clusters considered in figure 10 have a prolate structure. Interestingly, we find that the $α$ values increase with an increase of the number of Si atoms ($m$). In figure 10, we include the plots of $E_b$, $E_g$, and $Δq$ versus $m$ to further check the correlation between $α$ and $E_b$, $E_g$, and $Δq$, respectively. As shown in figure 10, both $E_b$ and $Δq$ are more available than $E_g$ for reflecting the size of $α$, that is, reverse and positive relationships with $α$ for $E_b$ and $Δq$, respectively, in agreement with the discussions above. Note that for the correlation between $α$ and both $E_b$ and $Δq$, we can only give the qualitative correlation for all the clusters considered, that is, the α values exhibit overall decreasing and increasing trends with increases in the $E_b$ and Δq values, respectively. We cannot give the generalized quantitative correlation expression for all the clusters considered.

## 4. Conclusions

We have theoretically investigated the size-, shape-, and composition-dependence of the polarizabilities for small $Si_mC_n$ ($m, n = 1–4$) clusters. The linear and prolate clusters with a delocalized π-electron framework have systematically larger polarizabilities than the compact ones, which is available for many small semiconductor clusters. For example, the prolate $Si_3$ ($C_{2v}$), $Si_4$ ($D_{2h}$), $Al_2P_2$ ($D_{2h}$), $Al_3P_3$ ($D_{3h}$), $Ga_2As_2$ ($D_{2h}$), and $Ga_4As$ ($C_{2v}$) clusters have a large polarizability per atom [1, 2, 8, 16, 24, 26]. Although both $E_g$ and $E_b$ are generally used to characterize the stability of the cluster, $E_b$ is more available than $E_g$ for reflecting the size of polarizability. For the isomers with the same $E_b$ values and different polarizabilities, over the factors studied in this work, the $Δq$ allows us to better explain the size-dependence of polarizabilities. The composition-dependence of $α$ suggests that the $α$ values of heteroatomic ($A_mB_n$) clusters should gradually converge to those of pure $A_m$ and $B_n$ clusters with an increase and decrease of the *m*/*n* value, respectively. We explicitly show by fitting a quantitative expression that a tunable polarizability can be obtained by adjusting the composition in clusters. Finally, to our knowledge, there has been no experimental polarizability for SiC cluster, therefore, for heteroatomic SiC clusters which simultaneously have these three dependences, our present results are useful references for future experiments on the polarizabilities of these clusters.

**Acknowledgement**

The authors are grateful for the project supported by Scientific Research Fund of Zhejiang Provincial Education Department [Y201119609] and the calculation support provided by the Foundation of Zhejiang Key Laboratory for Reactive Chemistry on Solid Surfaces.

Table 1. $\Delta q_x$, $\alpha_{xx}$, $\Delta q_y$, $\alpha_{yy}$, $\Delta q_z$, $\alpha_{zz}$, and $\Delta q$ of $Si_mC_n$ ($m$, $n$ = 1 – 4) clusters. "L", "P", and "C" indicate the linear, prolate, and compact clusters, respectively.

| Cluster | $\Delta q_x$ (e) | $\alpha_{xx}$ (Å$^3$) | $\Delta q_y$ (e) | $\alpha_{yy}$ (Å$^3$) | $\Delta q_z$ (e) | $\alpha_{zz}$ (Å$^3$) | $\Delta q$ (e/atom) |
|---|---|---|---|---|---|---|---|
| SiC (L, C$_{\infty v}$)[a] | 0.017 | 8.48 | 0.017 | 8.48 | 0.563 | 8.82 | 0.100 |
| SiC$_2$ (P, C$_{2v}$) | 0.017 | 5.49 | 0.286 | 6.87 | 0.452 | 7.90 | 0.084 |
| SiC$_2$ (L, C$_{\infty v}$) | 0.003 | 6.23 | 0.003 | 6.23 | 0.545 | 12.14 | 0.061 |
| SiC$_3$ (P, C$_{2v}$) (a) | 0.013 | 6.47 | 0.367 | 7.39 | 0.441 | 10.55 | 0.068 |
| SiC$_3$ (P, C$_{2v}$) (b) | 0.009 | 6.53 | 0.238 | 9.56 | 0.466 | 9.15 | 0.059 |
| SiC$_3$ (L, C$_{\infty v}$) | 0.013 | 7.21 | 0.013 | 7.21 | 0.932 | 21.95 | 0.080 |
| SiC$_4$ (P, C$_{2v}$) | 0.005 | 6.72 | 0.542 | 13.20 | 0.359 | 9.34 | 0.060 |
| SiC$_4$ (L, C$_{\infty v}$) | 0.011 | 7.79 | 0.011 | 7.79 | 0.895 | 29.69 | 0.061 |
| Si$_2$C (P, C$_{2v}$) | 0.000 | 9.02 | 0.709 | 18.24 | 0.263 | 9.05 | 0.108 |
| Si$_2$C (L, D$_{\infty h}$) | 0.003 | 9.26 | 0.003 | 9.27 | 0.722 | 19.10 | 0.081 |
| Si$_2$C$_2$ (P, D$_{2h}$) | 0.011 | 8.55 | 0.244 | 9.09 | 0.597 | 16.57 | 0.071 |
| Si$_2$C$_2$ (L, D$_{\infty h}$) | 0.023 | 10.13 | 0.023 | 10.13 | 1.330 | 36.69 | 0.115 |
| Si$_2$C$_3$ (P, C$_{2v}$) | 0.009 | 9.78 | 0.925 | 25.43 | 0.467 | 10.53 | 0.093 |
| Si$_2$C$_3$ (L, D$_{\infty h}$) | 0.011 | 10.59 | 0.011 | 10.59 | 1.201 | 42.65 | 0.082 |
| Si$_2$C$_4$ (L, D$_{\infty h}$) | 0.006 | 11.73 | 0.006 | 11.73 | 1.641 | 67.88 | 0.092 |
| Si$_2$C$_4$ (P, C$_s$) | 0.422 | 12.08 | 0.755 | 19.25 | 0.319 | 13.74 | 0.083 |
| Si$_3$C (P, C$_{2v}$) | 0.002 | 11.06 | 0.761 | 21.66 | 0.677 | 15.27 | 0.120 |
| Si$_3$C (P, C$_s$) | 0.276 | 11.95 | 0.789 | 17.59 | 0.719 | 20.52 | 0.149 |
| Si$_3$C$_2$ (P, C$_{2v}$) (a) | 0.036 | 12.12 | 1.267 | 31.64 | 0.417 | 13.96 | 0.115 |
| Si$_3$C$_2$ (C, C$_{2v}$) (b) | 0.371 | 14.07 | 0.618 | 19.27 | 0.851 | 17.98 | 0.123 |
| Si$_3$C$_3$ (P, C$_s$) (a) | 0.192 | 15.52 | 0.842 | 13.10 | 0.641 | 21.09 | 0.093 |
| Si$_3$C$_3$ (C, C$_s$) (b) | 0.636 | 12.78 | 0.470 | 20.76 | 0.692 | 22.96 | 0.100 |
| Si$_3$C$_4$ (P, C$_{2v}$) (a) | 0.008 | 21.21 | 0.874 | 13.63 | 0.706 | 26.50 | 0.076 |
| Si$_3$C$_4$ (C, C$_{2v}$) (b) | 0.714 | 32.91 | 0.317 | 19.52 | 0.989 | 13.07 | 0.096 |
| Si$_3$C$_4$ (P, C$_s$) | 0.992 | 12.63 | 0.758 | 28.17 | 0.050 | 21.04 | 0.086 |
| Si$_4$C (C, C$_{2v}$) | 0.813 | 23.56 | 0.621 | 21.35 | 0.715 | 18.02 | 0.143 |
| Si$_4$C (C, C$_{3v}$) | 0.923 | 21.61 | 0.880 | 21.61 | 0.707 | 18.43 | 0.167 |
| Si$_4$C$_2$ (C, C$_{2v}$) | 0.650 | 20.23 | 0.803 | 26.42 | 0.417 | 15.17 | 0.104 |
| Si$_4$C$_2$ (C, D$_{2d}$) | 0.714 | 22.21 | 0.714 | 22.21 | 0.438 | 14.47 | 0.104 |
| Si$_4$C$_3$ (P, C$_1$) | 1.184 | 33.66 | 1.010 | 26.44 | 0.466 | 16.57 | 0.127 |
| Si$_4$C$_3$ (P, C$_s$) | 0.929 | 26.12 | 1.292 | 35.21 | 0.466 | 15.16 | 0.128 |
| Si$_4$C$_4$ (P, C$_{2v}$) | 0.023 | 15.85 | 1.298 | 34.49 | 1.683 | 32.76 | 0.125 |
| Si$_4$C$_4$ (P, C$_s$) (a) | 0.924 | 28.70 | 1.310 | 35.99 | 0.026 | 15.95 | 0.094 |
| Si$_4$C$_4$ (C, C$_s$) (b) | 0.430 | 18.48 | 0.706 | 25.80 | 0.699 | 26.27 | 0.076 |
| Si$_4$C$_4$ (C, C$_s$) (c) | 0.722 | 22.18 | 0.924 | 30.16 | 0.574 | 22.49 | 0.092 |

[a] cluster shape and symmetry.

List of figure captions:

Figure 1. Geometries of Si$_m$C$_n$ ($m, n = 1 – 4$) clusters. "L", "P", and "C" indicate the linear, prolate, and compact clusters, respectively. [a] Electronic state and symmetry. [b] Binding energy (in eV/atom) and energy gap (in eV). [c] Cluster shape and $<\alpha>$ (in Å$^3$/atom).

Figure 2. Polarizabilities versus the cluster size ($N$) for Si$_m$C$_n$ ($m, n = 1 – 4$), Si$_m$ ($m = 1 – 8$), and C$_n$ (linear: $n = 1 – 8$, cyclic: $n = 4, 6,$ and 8) clusters. The polarizabilities of cyclic C$_n$ ($n = 4, 6,$ and 8) clusters are 1.39, 1.44, and 1.50 Å$^3$/atom, respectively. Dot line: Si bulk, Solid line: SiC bulk, Dash line: C bulk.

Figure 3. Polarizabilities versus the $E_g$ values for Si$_m$C$_n$ ($m, n = 1 – 4$) clusters.

Figure 4. Polarizabilities versus the $E_b$ values for Si$_m$C$_n$ ($m, n = 1 – 4$) clusters. The dot line indicates an approximate linear correlation between α and $E_b$.

Figure 5. Polarizabilities versus the $\Delta q$ values for Si$_m$C$_n$ ($m, n = 1 – 4$) clusters. The dot line indicates an approximate linear correlation between α and $\Delta q$.

Figure 6. Shape dependence of the $<\alpha>$ values for Si$_m$C$_n$ ($m, n = 1 – 4$) clusters.

Figure 7. (a) Polarizabilities, (b) Binding energies, (c) Energy gaps, and (d) $\Delta q$ versus the number of C atoms ($n$) in cluster with both the cluster shape and the number of Si atoms fixed. (L), (P), and (C) indicate the linear, prolate, and compact clusters, respectively. (+3.0) indicates 3.0 is added to the $<\alpha>$ values to clearly display the plots. Other factors have a similar meaning.

Figure 8. HOMO−1, HOMO, LUMO, and LUMO+1 of SiC$_2$, SiC$_4$, Si$_3$C$_2$, Si$_3$C$_3$, and Si$_4$C$_4$ clusters.

Figure 9. Polarizabilities versus the cluster composition (*i.e.*, *m*/*n* ratio) for Si$_m$C$_n$ (*m*, *n* = 1 – 4) clusters.

Figure 10. Polarizabilities (Å$^3$/atom), $E_b$ (eV/atom), $E_g$ (eV), and $\Delta q$ (e/atom) versus the number of Si atoms (*m*) in cluster with the number of C atoms fixed. (+1.0) indicates 1.0 is added to the $E_b$ values to clearly display the plots. Other factors have a similar meaning.

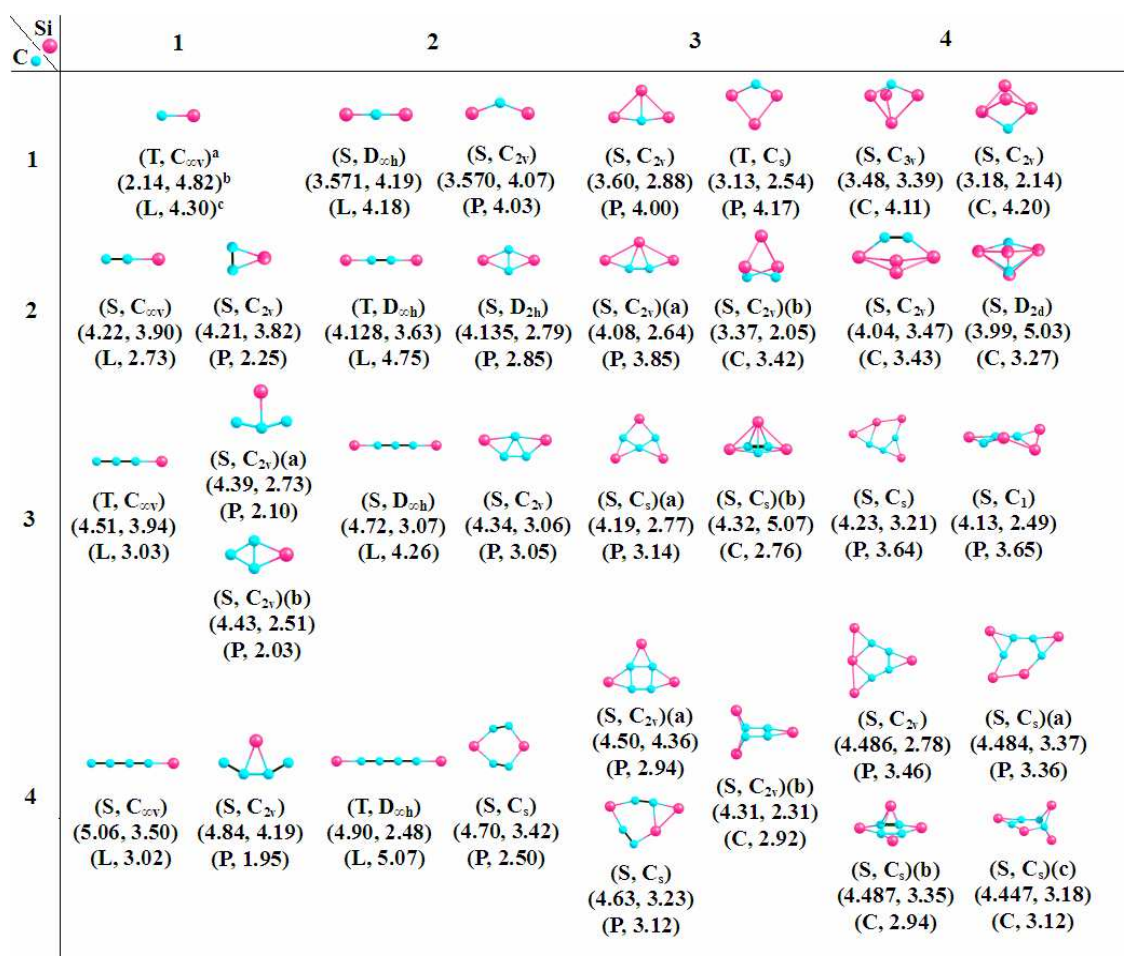

Figure 1. Geometries of Si$_m$C$_n$ (*m*, *n* = 1 – 4) clusters. "L", "P", and "C" indicate the linear, prolate, and compact clusters, respectively.
[a] Electronic state and symmetry.
[b] Binding energy (in eV/atom) and energy gap (in eV).
[c] Cluster shape and <*α*> (in Å$^3$/atom).

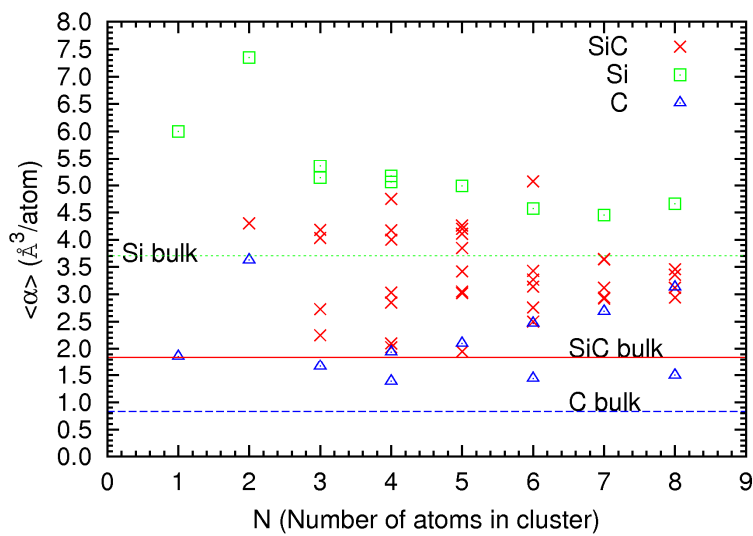

Figure 2. Polarizabilities versus the cluster size (*N*) for $Si_mC_n$ (*m*, *n* = 1 – 4), $Si_m$ (*m* = 1 – 8), and $C_n$ (linear: *n* = 1 – 8, cyclic: *n* = 4, 6, and 8) clusters. The polarizabilities of cyclic $C_n$ (*n* = 4, 6, and 8) clusters are 1.39, 1.44, and 1.50 Å$^3$/atom, respectively. Dot line: Si bulk, Solid line: SiC bulk, Dash line: C bulk.

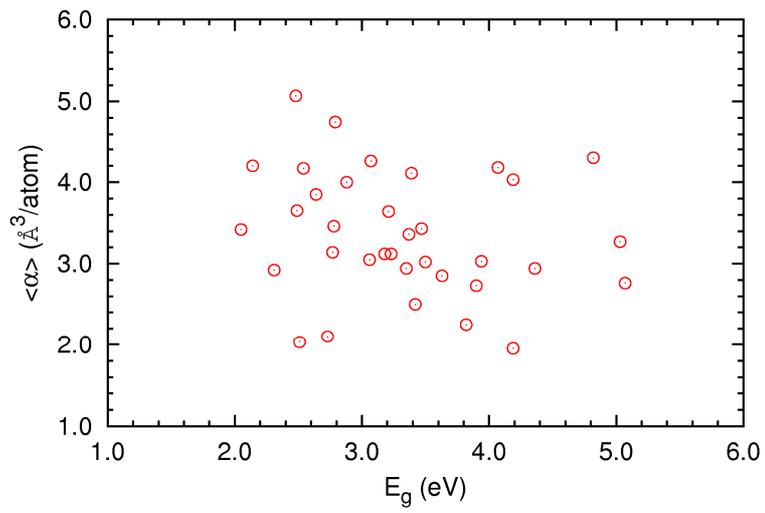

Figure 3. Polarizabilities versus the $E_g$ values for $Si_mC_n$ ($m$, $n$ = 1 – 4) clusters.

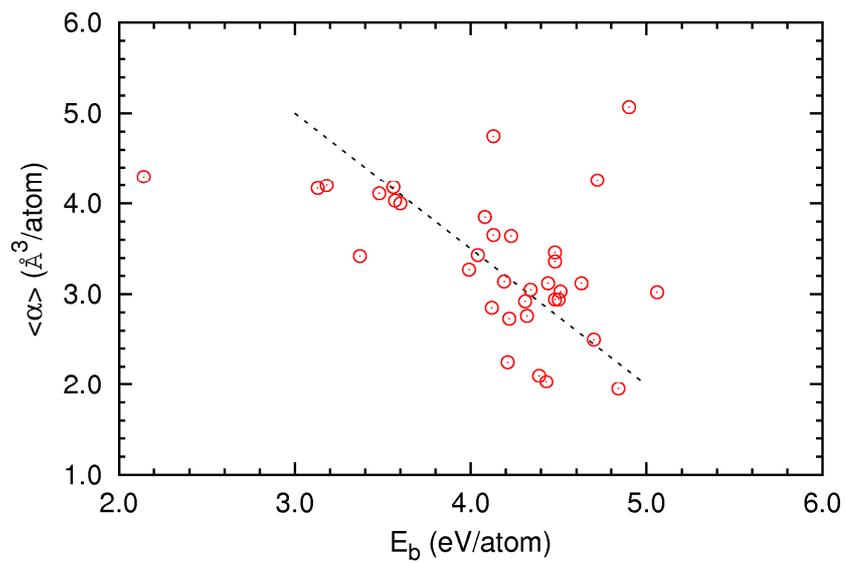

Figure 4. Polarizabilities versus the $E_b$ values for $Si_mC_n$ ($m$, $n$ = 1 – 4) clusters. The dot line indicates an approximate linear correlation between α and $E_b$.

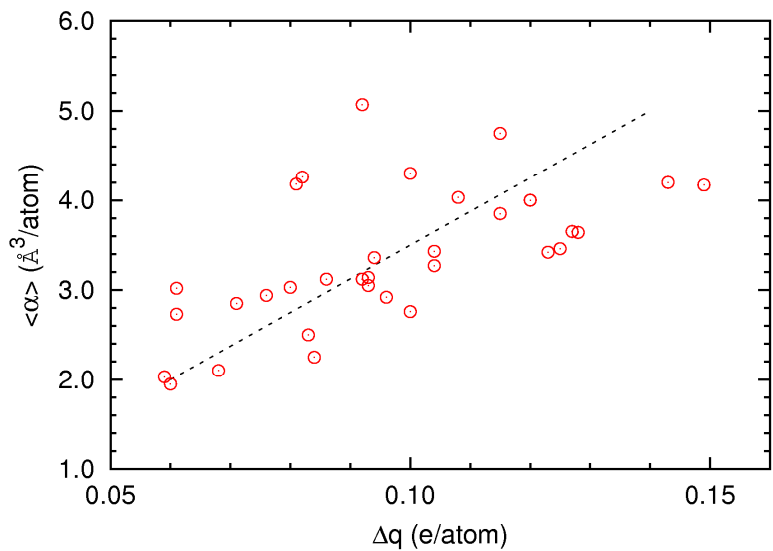

Figure 5. Polarizabilities versus the Δ*q* values for Si$_m$C$_n$ (*m*, *n* = 1 – 4) clusters. The dot line indicates an approximate linear correlation between α and Δ*q*.

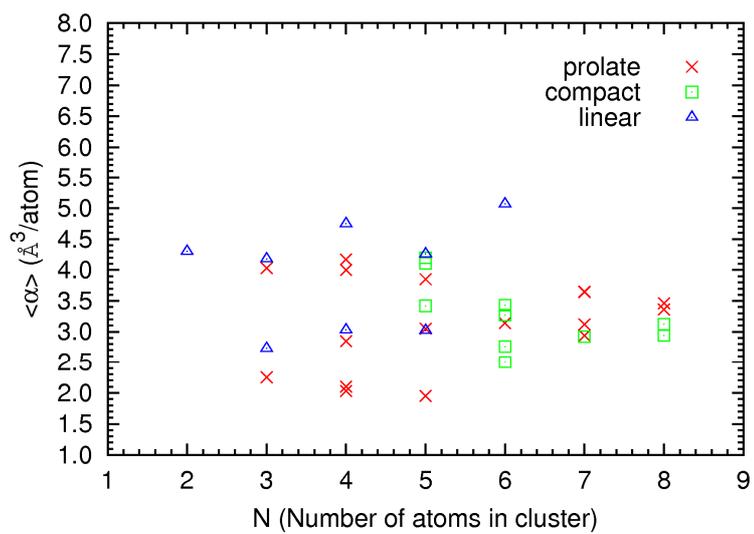

Figure 6. Shape dependence of the $\langle\alpha\rangle$ values for $Si_mC_n$ ($m, n = 1 – 4$) clusters.

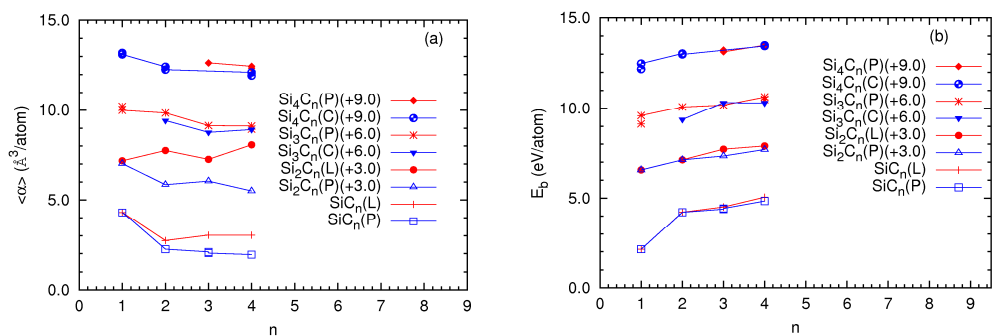

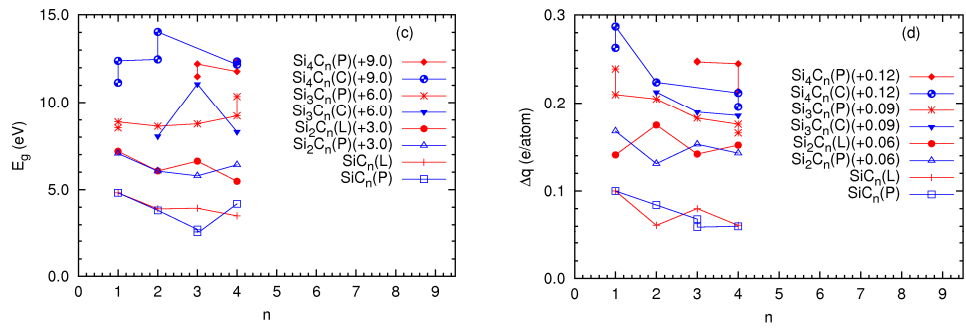

Figure 7. (a) Polarizabilities, (b) Binding energies, (c) Energy gaps, and (d) Δ*q* versus the number of C atoms (*n*) in cluster with both the cluster shape and the number of Si atoms fixed. (L), (P), and (C) indicate the linear, prolate, and compact clusters, respectively. (+3.0) indicates 3.0 is added to the <*α*> values to clearly display the plots. Other factors have a similar meaning.

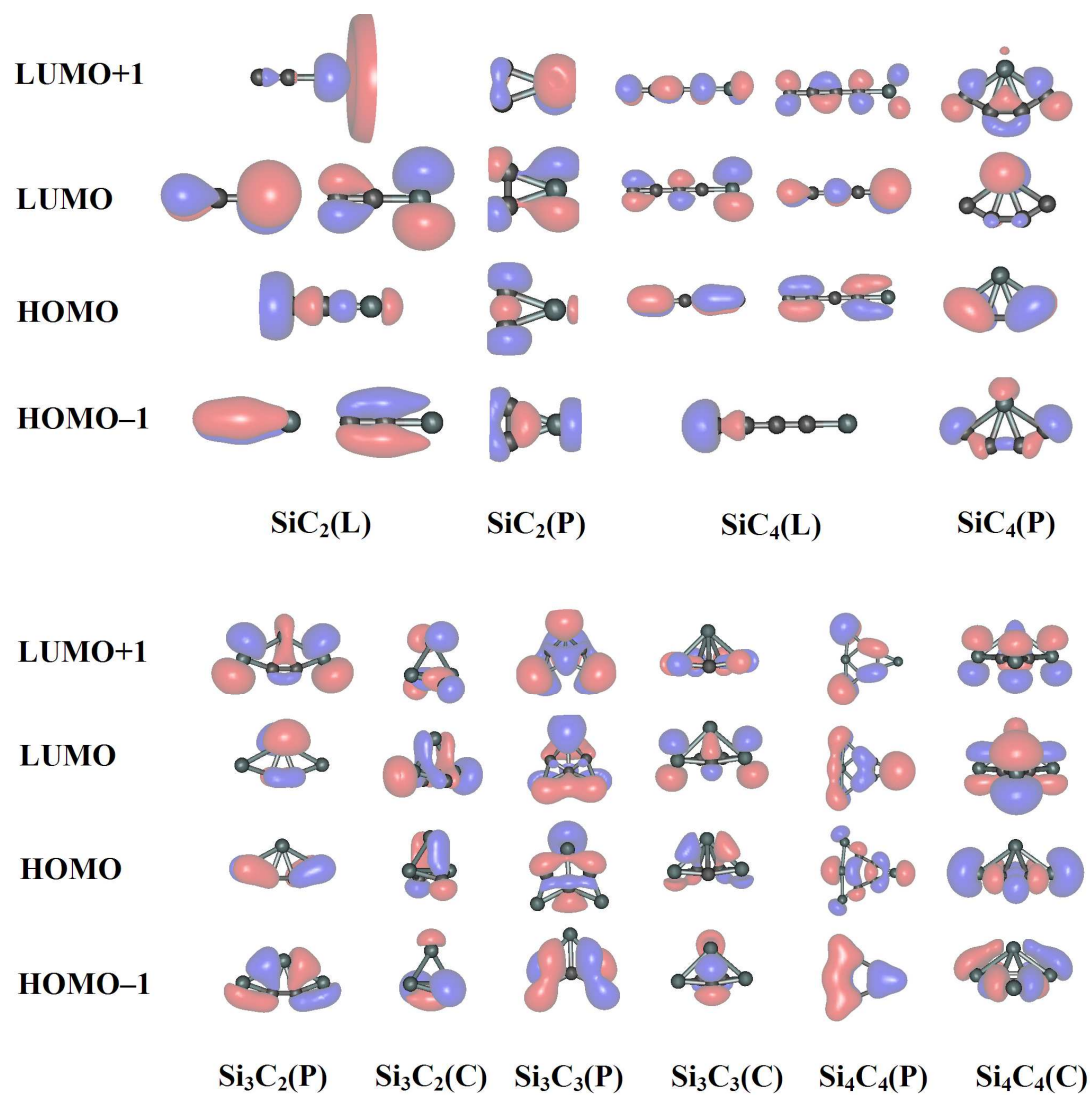

Figure 8. HOMO−1, HOMO, LUMO, and LUMO+1 of $SiC_2$, $SiC_4$, $Si_3C_2$, $Si_3C_3$, and $Si_4C_4$ clusters.

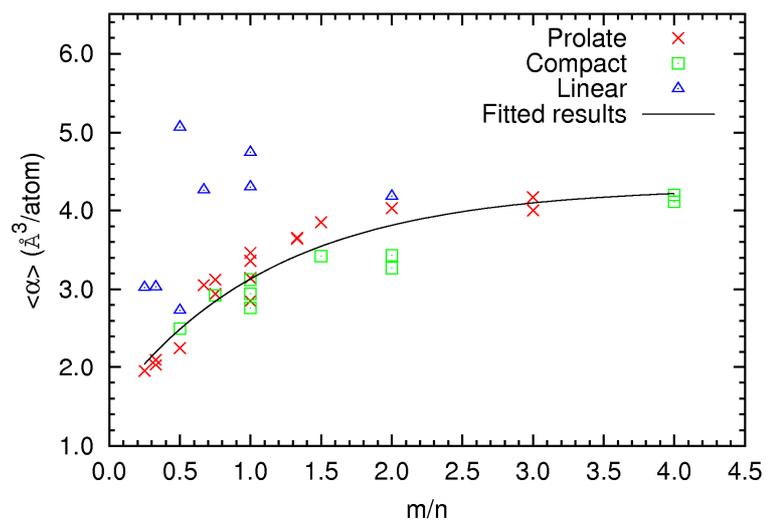

Figure 9. Polarizabilities versus the cluster composition (*i.e.*, *m/n* ratio) for Si$_m$C$_n$ (*m*, *n* = 1 – 4) clusters.

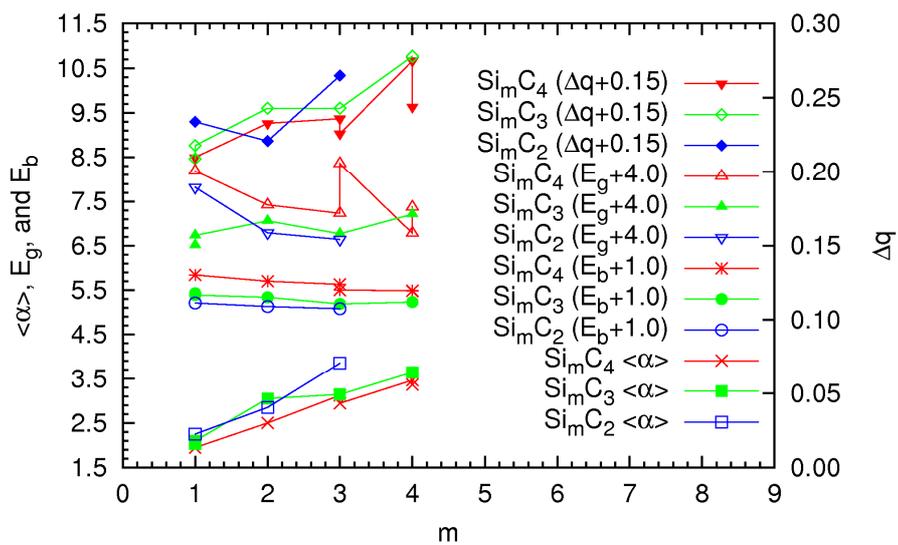

Figure 10. Polarizabilities (Å³/atom), $E_b$ (eV/atom), $E_g$ (eV), and $\Delta q$ (e/atom) versus the number of Si atoms (m) in cluster with the number of C atoms fixed. (+1.0) indicates 1.0 is added to the $E_b$ values to clearly display the plots. Other factors have a similar meaning.